\documentclass[11pt]{article}
\usepackage[utf8]{inputenc}
\usepackage{float}
\usepackage{setspace}
\usepackage[margin=1in]{geometry}

\usepackage[open,openlevel=1]{bookmark}

\usepackage{authblk}
\usepackage{natbib}
\usepackage{graphics}
\usepackage[table]{xcolor}
\usepackage{booktabs} 
\usepackage{comment}
\usepackage{amssymb}
\usepackage{graphicx}

\usepackage[centertags,reqno]{amsmath}
\usepackage{mathtools}
\usepackage[thmmarks,amsmath]{ntheorem}

\theoremnumbering{arabic}
\theoremstyle{break}
\theoremsymbol{}
\theoremheaderfont{\normalfont\bfseries}
\theorembodyfont{\upshape}

\theoremstyle{nonumberbreak}

\usepackage{adjustbox}

\usepackage{caption}
\usepackage{subcaption}

\usepackage[normalem]{ulem}
\useunder{\uline}{\ul}{}

\usepackage{bm}

\providecommand{\keywords}[1]
{
  \small	
   \quad \textbf{Keywords:} #1
}

\numberwithin{equation}{section}
\numberwithin{table}{section}
\numberwithin{figure}{section}

\begin{document}

\title{Effects of model misspecification on small area estimators}

\author{Yuting Chen, Partha Lahiri, and Nicola Salvati \footnote{Yuting Chen, Partha Lahiri, and Nicola Salvati, University of Maryland College Park, University of Maryland College Park, and Università di Pisa, Email: ychen215@umd.edu}}



\date{\vspace{-5ex}}

\maketitle

\begin{abstract}
Nested error regression models are commonly used to incorporate unit specific auxiliary variables to improve small area estimates. When the mean structure of the model is misspecified, the design-based mean squared prediction error (MSPE) of Empirical Best Linear Unbiased Predictors (EBLUP) generally increases. The Observed Best Prediction (OBP) method has been proposed with the intent to improve on the design-based MSPE over EBLUP. 
In this paper, we conduct a Monte Carlo simulation experiments to understand the effect of misspsecification of mean structures on different small area estimators. Our findings suggest that the OBP using unit-level auxiliary variables does not outperform the EBLUP in terms of design-based MSPE, unless the number of small areas $m$ is extremely large. Conversely, the performance of OBP significantly improves when area-level auxiliary variables are employed. This paper includes both analytical and numerical evidence to demonstrate these observations, providing practical insights for addressing model misspecification in SAE.
\end{abstract}

\newpage

\section{Introduction}
Direct survey-weighted estimates \citep[e.g.,][]{cochran1977sampling} are routinely used for producing a wide range of socio-economic, environment, health, and other official statistics for national and large sub-national areas. When the targets are smaller geographical areas, the issue of having a limited or no sample can render direct estimates unreliable. Therefore, small area estimation (SAE) has garnered increasing attention in recent decades, as it is used to leverage strength from related data sources such as census, administrative and geo-spatial data through statistical modeling. For a comprehensive review of various SAE methods and models, we refer to \cite{rao2015small}.

\cite{battese1988error} introduced a nested error regression (NER) model that links the response variable to the auxiliary variables at the unit level:
\begin{eqnarray}\label{unitlevel}
y_{ij} = \bm{x}_{ij}^\prime\bm \beta + v_i + e_{ij}, \quad i = 1, \cdots, m, \quad j = 1, \cdots, N_i,
\label{unitmod}
\end{eqnarray}
where $m$ is the number of areas;  $N_i$ is known population size of area $i$;  $v_i$'s are area-specific random effect; $e_{ij}$'s are sampling errors. We assume  that $\{v_i,\;i=1,\cdots,m\}$ and  $\{e_{ij},\;i=1,\cdots,m; j=1,\cdots,N_i\} $ are independent with $v_i\sim N(0, \sigma^2_v)$ and $e_{ij}\sim N(0, \sigma^2_e)$. The regression coefficients $\bm \beta$ and the variance components $\sigma^2_v$ and $\sigma^2_e$ are generally unknown. In this section, for simplicity of exposition, we assume  that a simple random  sample (SRS) of size $n_i$ is selected for small area $i$.


In many applications, unit-level auxiliary variables are not available or are outdated. In such cases, models like the following special case of the NER model \citep{newhouse2022small}, often referred to as the unit-context model, are used:
\begin{eqnarray}
y_{ij} = \bar{\bm X}_{i}^\prime \bm\beta + v_i + e_{ij}, \quad i = 1, \cdots, m, \quad j = 1, \cdots, N_i,
\label{ucmod}
\end{eqnarray}
where $\bar{\bm X}_{i}=N_i^{-1}\sum_{j=1}^{N_i}\bm{x}_{ij}$ is  a vector of known population means of auxiliary variables for area $i$. The same assumptions on the random effects $v_i$ and sampling errors $e_{ij}$ apply.

Area-level model, introduced by \cite{fay1979estimates}, can be another possibility to handle lack of good unit-specific auxiliary variables. An area level model that corresponds to the unit level model (\ref{unitlevel}) is given by:
\begin{eqnarray}
\bar{y}_{i} = \bar{\bm X}_{i}^\prime \bm \beta + v_i + \bar{e}_{i}, \quad i = 1, \cdots, m,
\label{fhmod}
\end{eqnarray}
where $\bar{y}_i$ is the unweighted sample mean;  $v_i$'s are area-specific random effect; $\bar{e}_{i}$'s are sampling errors. It is assumed that $v_i$'s and  $\bar{e}_{i}$'s are independent with $v_i\stackrel{iid}\sim N(0, \sigma^2_v)$, $\bar{e}_{i}\stackrel{iid}\sim N(0, D_i)$, where $D_i$ is known sampling variance of $\bar y_i$. In practice, $D_i$'s are estimated using a smoothing technique such as the one given in \cite{otto1995sampling}.  In this case, one may propose a smoothed estimator of $D_i$ as $s^2/n_i$, where $s^2$ is the pooled sample variance using data from all $m$ areas.

The Best Predictor (BP) of a mixed effect in a linear mixed model is simply the conditional mean of the mixed effect given the data.   The explicit formula for the BP of small area mean can be explicitly obtained for both unit level and area level models. 
As in \cite{jiang2015observed}, we assumed normality of both model and sampling errors.  However, note that normality is not required for obtaining the BP of small area means.  For example, the BP can be obtained under the assumption that the conditional means of the random effects given the data is  a linear function of the sample observations; see, e.g., \cite{ghosh1987robust}. The Best Linear Unbiased Predictor (BLUP) is obtained from the Best Predictor (BP) under the assumed linear mixed model when the unknown regression coefficients in the BP is replaced by the weighted least square estimators. Estimated BLUP (EBLUP) is obtained when the unknown variance components of the model are replaced by standard estimators (e.g., REML). The BLUP and EBLUP can be viewed as an estimated BP  (EBP) of the mixed effects.

\cite{jiang2011best} proposed Observed Best Predictors (OBP) for small area means under a linear mixed model in an attempt to reduce the effects of misspecification of the mean structure in the assumed model.  We note that their OBP can be motivated as EBP when best predictive estimators (BPE) of model parameters, which minimize the observed mean squared prediction error, are used in place of standard model parameter estimators. In the context of area level models, \cite{jiang2011best} used both theoretical and empirical studies to demonstrate that OBP outperforms EBLUP in terms of design-based MPSE when the underlying linear mixed model is misspecified. Subsequently, \cite{jiang2015observed} considered OBP for a nested error regression model, where both the mean function and variance components are misspecified. Their simulations indicated that OBP may perform significantly better than EBLUP in terms of both overall and area-specific design-based MSPE. However, our simulation studies indicate that OBP using unit-level auxiliary variables (OBP-UNIT) does not  outperform EBLUP unless the number of areas $m$ is extremely large, which is unusual in small area estimation. We found that using area-level auxiliary variables, such as in the unit-context or area-level models, one can improve the performance of OBP.

In this paper, we investigate the effects of misspecified mean function and variance components on the predictive performances of existing small area estimators. We also provide both analytical and numerical evidence explaining why OBP-UNIT may underperform in terms of design-based MSPE, and how the use of area-level auxiliary variables can enhance its effectiveness. The rest of the paper is organized as follows. In Section \ref{sec:2}, we  provide further details on OBP methodology and the reasoning why the OBP with area-level auxiliary variables outperforms the OBP with unit level auxiliary variables. In Section \ref{sec:3}
we  present numerical studies and the evaluation results in terms of both overall and area-specific design-based MSPE. Section \ref{sec:4} offers conclusions and practical guidelines for implementing OBP in the presence of model misspecification.

\section{Analytical comparison between OBP with unit-level auxiliary variables and OBP with area-level auxiliary variables }
\label{sec:2}
Let $\bm{\psi} = (\bm{\beta}^\prime,\sigma^2_u, \sigma^2_e)$ represent the vector of model parameters of the nested error regression model, and let $\tilde{\bm{\theta}}(\bm{\psi}) = [\tilde{\theta}_i(\bm{\psi})]_{1\leq i \leq m}$ denote the vector of BPs for the true small area means $\bm{\theta} = [\theta_i]_{1\leq i \leq m}$. The design-based  MSPE of $\tilde{\bm{\theta}}(\bm{\psi})$ is defined as
\begin{equation}
    \text{MSPE}(\tilde{\bm{\theta}}(\bm{\psi})) = \text{E}(|\tilde{\bm{\theta}}(\bm{\psi}) - \bm{\theta}|^2) = \sum_{i=1}^m\text{E}(\tilde{\theta}_i(\bm{\psi}) - \theta_i)^2,
    \label{MSPE=EQ11}
\end{equation}
where the expectation is with respect to the sample design.
Following \cite{jiang2011best} and \cite{jiang2015observed}, the MSPE in \eqref{MSPE=EQ11} has an alternative expression, which is a key idea of the OBP. Namely, it leads to the fundamental equation of the OBP, 
\begin{equation}
            \text{MSPE}(\tilde{\bm{\theta}}(\bm{\psi})) = \text{E}\{Q(\bm{\psi})\},
            \label{MSPE=EQ12}
\end{equation}
and $Q(\cdot)$ is called the observed MSPE function. Under the nested error regression model, the $Q$ function can be expressed as:
\begin{equation}
\begin{split}
 Q =& \bm{\beta}^\prime(\bar{\bm{X}} -\bm{G}\bar{\bm x})^\prime(\bar{\bm{X}} -\bm{G}\bar{\bm x})\bm{\beta}
- 2\bar{\bm y}^\prime\{(\bm{I}_m - 2\bm{G})\bar{\bm X} + \bm{G}^2\bar{\bm x}\}\bm{\beta} \\
      &+ \bar{\bm y}^\prime \bm{G}^2 \bar{\bm y} + \bm{1}_m^\prime(\bm{I}_m - 2\bm{G})\hat{\bm \mu}^2,\\
\end{split}
\end{equation}
where $\bar{\bm X} = (\bar{\bm X}_i^\prime)_{1\leq i \leq m}$, $\bar{\bm x} = (\bar{\bm x}_i^\prime)_{1\leq i \leq m}$, $\bar{\bm x}_i$ being the vector of  unweighted sample means of the auxiliary variables for area $i$, $\bar{\bm y} = (\bar y_i)_{1\leq i \leq m}$, $\bm{G} = \text{diag}\{n_i/N_i + (1 - n_i/N_i)n_i\sigma^2_v/(n_i\sigma^2_v+\sigma^2_e), 1\leq i \leq m\}$ and $\hat{\bm\mu}^2 = (\hat{\mu}^2_i)_{1\leq i\leq m}$is the design-unbiased estimator of $(\bar{Y}^2_{i})_{1\leq i \leq m}$.
The BPE of $\bm{\psi}$, denoted as $\hat{\bm{\psi}}$, is the minimizer of $Q(\bm{\psi})$ with respect to $\bm{\psi}$.


To understand why OBP performs optimally when area-level auxiliary variables are used, it is instructive to consider the case when all the unit-level model parameters $\bm{\psi}$ are known.   
Using the facts that  $\text{E}(\bar{\bm x} - \bar{\bm X}) = 0$ and $\text{E}\{\bm{\bar{y}}^\prime\bm{G}^2(\bm{\bar x} - \bm{\bar{X}})\bm{\beta}\} = \text{cov}(\bar{\bm y}^\prime\bm{G}^2, \bm{\beta}^\prime\bar{\bm x}^\prime)$, we get
\begin{equation}
  \label{EQ*}
  \begin{split}
      \text{E}Q 
       = & \text{E}\{\bm{\beta}^\prime(\bm{\bar x} - \bm{\bar{X}})^\prime\bm{G}^2(\bm{\bar x} - \bm{\bar{X}})\bm{\beta}\} + \cdots
  \end{split}
\end{equation}
where $\cdots$ are the terms involving finite population parameters  and known model parameters. Since $\bm{G}^2$ is positive definite, the expression is minimized at $\bar{\bm x} - \bar{\bm X} = 0$, for any fixed $\bm{\beta}$, indicating that the use of area-level auxiliary variables leads to optimal OBP performance. 

\section{Simulations}\label{sec:3}
The more detailed derivations of OBP procedure for both area-level and unit-level models can be found in the supplementary materials of \cite{jiang2011best}. Also, since unit-context model is a special type of unit-level models, we can directly derive OBP under a unit-context model by replacing $\bm x_{ij}$ by $\Bar{\bm X}_i$.

The simulation settings are similar to \cite{jiang2015observed}. For simplicity, we consider a case of a single auxiliary variable that is assumed to be linearly associated with the response variable $y_{ij}$ through the following model:
\begin{eqnarray}
y_{ij} = \beta_1 x_{ij} + v_i + e_{ij}\quad
i=1,\dots, m, j=1,\dots, N_i,
\label{sim_mod1}
\end{eqnarray}
where $x_{ij}$'s are known values of an auxiliary variable for the $j$th unit of the $i$th area; $\beta_1$ is an unknown regression coefficient; $v_i$, $e_{ij}$ are the same as in \eqref{unitmod}. We assume that $x_{ij}$'s are not all the same in an area.  In the present context, \eqref{ucmod} becomes:
\begin{eqnarray}
y_{ij} = \beta_1 \bar X_{i} + v_i + e_{ij}\quad
i=1,\dots, m, j=1,\dots, N_i.
\label{sim_mod1_uc}
\end{eqnarray}
And the corresponding area-level model is given by:
\begin{eqnarray}
 \bar{y}_{i} = \beta_1 \bar X_{i} + v_i + \bar{e}_{i}\quad
i=1,\dots, m.
\label{sim_mod1_fh}
\end{eqnarray}

For the simulations set up, we draw simple random sampling without replacement (SRSWOR) samples from the population of each small areas and consider the following estimators of the small area means throughout the rest of this section:
\begin{itemize}
    \item[\bf (A)] Direct estimator (sample mean),
    \item[\bf (B)] OBP under the assumed unit-context model \eqref{sim_mod1_uc} (OBP-UC),
    \item[\bf (C)] OBP under the basic Fay-Herriot (area-level) model \eqref{sim_mod1_fh} with smoothed direct variance estimates: $\hat D_i =  s^2/n_i$, $s^2 = (n-1)^{-1}\sum_{i=1}^m\sum_{i=1}^{n_i}(y_{ij}-\bar y)^2$ and $n = \sum_{i = 1}^{m}n_i$, 
    \item[\bf (D)] OBP under the assumed unit-level model\eqref{sim_mod1} (OBP-UNIT), 
    \item[\bf (E)] EBLUP under the assumed unit-level model \eqref{sim_mod1} (EBLUP-UNIT),
    \item[\bf (F)] EBLUP under the assumed unit-context model \eqref{sim_mod1_uc} (EBLUP-UC).
\end{itemize}

To introduce model misspecifications, we first generate $y$ for the finite population from the following superpopulation heteroscedastic nested-error regression model: 
\begin{eqnarray}
y_{ij} = b+v_i+e_{ij}\quad
i=1,\dots, m, j=1,\dots, N_i.
\label{sim_gen1}
\end{eqnarray}
The population and sample sizes are the same for all areas and are fixed at $N_i = 1000$ and $n_i = 4$, respectively. We consider two different values of $b$: $b = 5, 10$, and three values of the number of small areas: $m = 40, 100$ or $400$ with $v_i$ generated from the normal distribution $N(0, 1)$, and $e_{ij}$ generated from the normal distribution $N(0,\sigma^2_{ei})$, where $\sigma^2_{ei}$ is independently generated from the gamma distribution $\Gamma(3,0.5)$. In each case, $x$ for the finite population is generated from a log-normal superpopulation  distribution with a mean of 1 and a standard deviation of 0.5.

Each scenario is independently simulated $K = 1000$ times. The performance of the estimators (A)-(F), under the above simulation setups, are assessed in terms of both overall and area-specific design-based MSPEs.  The area-specific design-based MSPE is defined as $\text{MSPE}(\hat{\bar{Y}}_i) = \text{E}(\hat{\bar{Y}}_i - \bar{Y}_i)^2$, where $\bar{Y}_i = N_i^{-1}\sum_{i=1}^{N_i}y_{ij}$ is the true small area mean, and $\hat{\bar{Y}}_i$ is the predicted value for $i$th area, either by OBP or EBLUP and E is with respect to the sample design. In Monte Carlo simulations, MSPE for area $i$ and overall MSPE are approximated by:
\begin{eqnarray}
{\text{MSPE}_i} \approx \frac{1}{K}\sum_{k =1}^{K}(\hat{\bar{Y}}_i^{(k)} - \bar{Y}_i^{(k)})^2,
\label{mspe_area}
\end{eqnarray}
\begin{eqnarray}
{\text{MSPE}}
\approx \frac{1}{m}\sum_{i=1}^{m}\left[\frac{1}{K}\sum_{k =1}^{K}(\hat{\bar{Y}}_i^{(k)} - \bar{Y}_i^{(k)})^2\right],
\label{mspe_overall}
\end{eqnarray}
respectively, where 
$\bar{Y}_i^{(k)}$ and  
$\hat{\bar{Y}}_i^{(k)}$ are the true mean and estimated mean for area $i$ 
in the $k^{th}$ simulation run, respectively. 

\begin{table}[h]
\centering
\caption{Overall simulated design-based MSPE when the finite population is generated from the heteroscedastic NER  model \eqref{sim_gen1}}
\begin{tabular}{c|cc c cc c}
    \toprule[1pt]
         $(m, b)$ & DIRECT & OBP-UC & OBP-FH & OBP-UNIT & EBLUP-UNIT & EBLUP-UC\\
    \midrule[0.75pt]
        (40, 5) &  1.475   & 0.689   & 0.698   &1.775  & 1.474 &0.687\\
        (100, 5)&  1.488   & 0.638   & 0.657   & 1.409 & 1.495 &0.638\\
        (400, 5)&  1.493   & 0.614   & 0.637   & 1.352 & 1.508 &0.613\\
        (40, 10)&  1.475   & 0.696   & 0.705   &3.919  & 1.522 & 0.694 \\ 
        (100, 10)& 1.488   & 0.646   & 0.664   &2.364  & 1.513 & 0.645\\ 
        (400, 10)& 1.493   & 0.621   & 0.644   & 1.711 & 1.510 & 0.621\\
    \bottomrule[1pt]
\end{tabular}
\label{tab:table1}  
\end{table} 

Table \ref{tab:table1} reports the simulated overall design-based MSPE for the various simulation conditions and estimators when the finite population is generated based on the true underlying model \eqref{sim_gen1}. The direct estimator is unaffected by model misspecification as it is not model-based. EBLUP-UNIT behavior is similar to direct because it automatically assigns more weight to the sample mean when the model is weak. While OBP-UNIT was designed to handle model misspecification, it surprisingly performs poorly, even worse than the direct estimator and EBLUP-UNIT when $b = 10$. 
OBP-UC and EBLUP-UC show similar and better performance than EBLUP-UNIT. The reason why EBLUP-UC also performs well could be that, given the large population size $N_i$ and the fact that the population values come from a common distribution across all areas, the population means $\bar{X}_i$ are roughly equal across areas. As a result, the unit-context model and the area-level model approximate the true model closely. Consequently, predictors using area-level auxiliary variables tend to outperform those using unit-level auxiliary variables. Overall, our findings suggest that in instances of significant model misspecification, OBP-UNIT may not be an effective choice.

\begin{figure}[h]
\centering
\caption{Distributions of area-specific simulated design-based MSPEs when the finite population is generated from the heteroscedastic NER  model \eqref{sim_gen1}}
\includegraphics[width=0.9\linewidth, height=13cm]{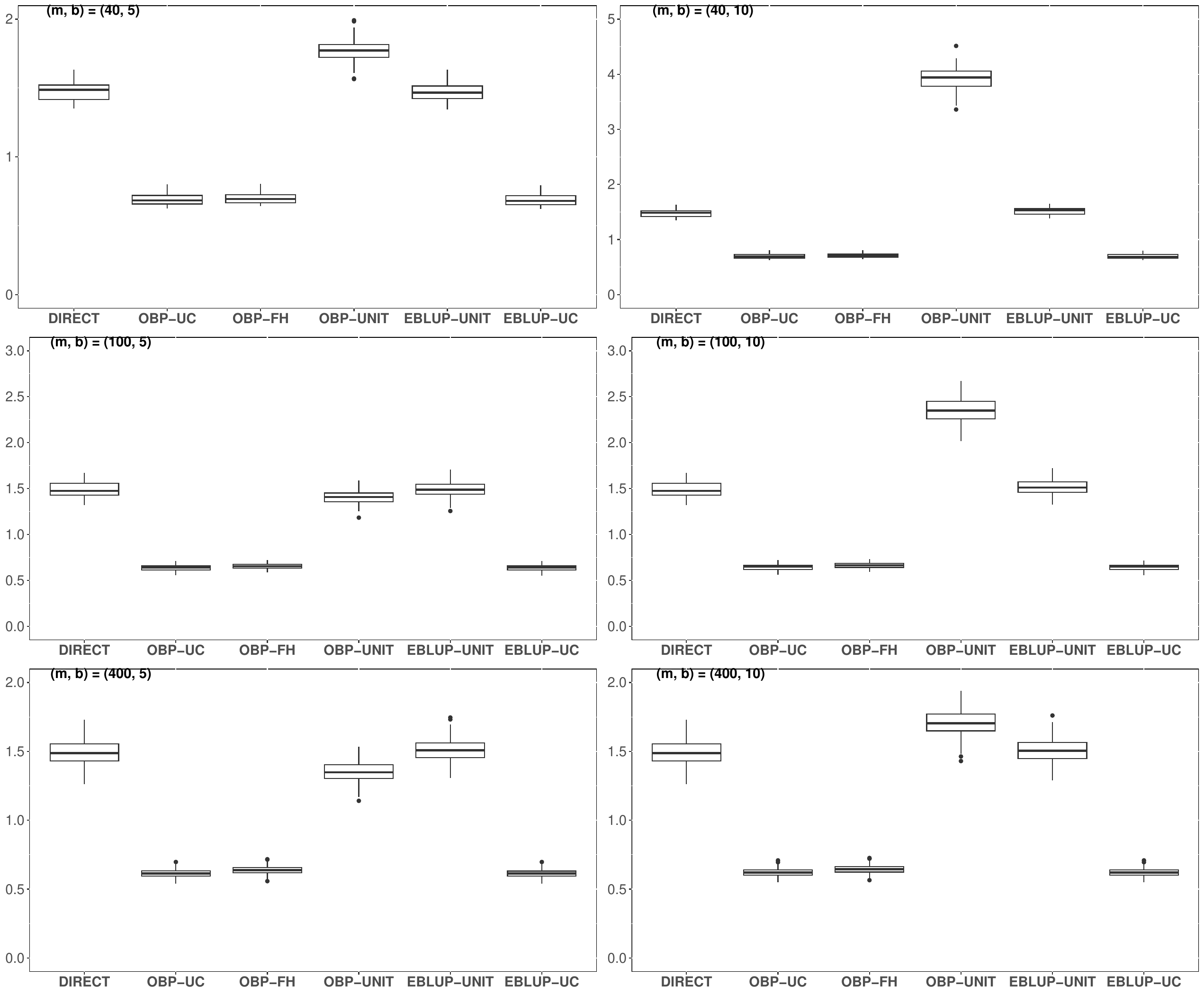} 

\label{fig:fig1}
\end{figure}

As for the area-specific MSPEs, we utilize boxplots to display the distributions of the area-specific design-based MSPEs associated with all the estimators. See Figure \ref{fig:fig1}. The boxplot of OBP-UNIT shows much larger median MSPE and variability than those of OBP-UC and OBP-FH and in some cases worse than direct and EBLUP-UNIT.  OBP-UC shows slightly larger variability than OBP-FH. It might be because that in contrast to area-level models, unit-context models incorporate the uncertainty resulting from the estimation of model parameters, such as sampling variances. 

As discussed in \cite{jiang2015observed}, the simulation conditions above might be a little extreme in which the assumed models are completely different from the true underlying model. This motivates us to consider some moderate cases where assumed model is partially correct compared to the true model. Keeping the same assumed models \eqref{sim_mod1} - \eqref{sim_mod1_fh} we generate the finite population for $y$ from the underlying superpopulation model:
\begin{eqnarray}
y_{ij} = b_0 + b_1 x_{ij} + v_i + e_{ij}\quad
i=1,\dots, m, j=1,\dots, N_i
\label{sim_mod2}
\end{eqnarray}
where $b_0 = 10$, $b_1 = 5$. Note that the slope in \eqref{sim_mod2} is not zero and it matches the linear relationship part of the assumed model \eqref{sim_mod1}. But still, we have slight misspecifications in the sense that the true model here also has a nonzero intercept. In addition, $e_{ij}$ is generated from the same normal distribution as described previously, and so we have the issue of heteroscedasticity as well. Three different values of $m$ are considered: $m = 40, 100$ or $400$, and $v_i$ is generated from normal distribution $N(0, 1)$. 

The results based on $K=1000$ simulations are displayed in Table \ref{tab:table2}. As expected, in this case EBLUP-UNIT outperforms the direct estimator since the assumed model is closer to the true model and EBLUP can borrow strengths from other related areas. While OBP-UNIT was expected to perform comparably to EBLUP-UNIT in this scenario, our simulation results indicate that OBP-UNIT performs actually worse than EBLUP-UNIT in terms of the simulated MSPE. When $m=40$, the simulated MSPE for OBP-UNIT is similar to that of the DIRECT estimator, indicating poor performance. In contrast, both OBP-UC and EBLUP-UC show similar results, performing slightly better than EBLUP-UNIT when $m=100$ and $m=400$. OBP-FH demonstrates stable performance and performs better than OBP-UC and EBLUP-UC when $m=40$.

The results in Tables \eqref{tab:table1} and \eqref{tab:table2} indicates that OBP-UNIT may not  effectively reduce the impact of model misspecification, compared with EBLUP-UNIT. This surprising outcome prompted further investigation, which we explore in the following Remarks 1 and 2.
\begin{table}[h!]
    \centering
        \caption{Overall simulated design-based MSPE when the finite population is generated from the heteroscedastic NER  model \eqref{sim_mod2}}
        \label{tab:table2} 
        \begin{tabular}{c|cc cccccc}
    \toprule[1pt]
        $m$&  DIRECT& OBP-UC &OBP-FH  & OBP-UNIT  & EBLUP-UNIT & EBLUP-UC \\
        \midrule[0.75pt]
        40 &  18.243  & 1.798  & 1.588  &18.230 & 1.532 & 1.872 \\
        100 & 18.339  & 1.341  & 1.248  &7.958 & 1.514 & 1.358\\
        400&  18.422  & 1.085  & 1.059  &3.122 & 1.509 & 1.087\\ 
         \bottomrule[1pt]
    \end{tabular}
\end{table}

\subsection*{Remark 1: MSPE of OBP-UNIT vs  $|\bar{\bm X} - \bar{\bm x}|$}
\label{remark1}

To further investigate the relationship between the performance of OBP-UNIT and the absolute difference between sample means and population means, we conducted a numerical study using the following Nested Error Regression (NER) superpopulation model: 
\begin{eqnarray}
 y_{ij} = 10 + v_i + e_{ij}\quad
i=1,\dots, 50,\; j=1,\dots, 1000.
\label{num_mod}
\end{eqnarray}
The generative processes for the finite population for  $x$, $v_i$, and $e_{ij}$ remain consistent with the aforementioned model. Next, we drew SRS samples of size $n_i=4$ for each small areas. 

In a simulation setting, when the number of units within a specific area is relatively small, it is often impossible to have sample mean $\bar{\bm x}$ equal or even close to population mean $\bar{\bm X}$. To explore how deviations of sample means $\bar{\bm x}$ from population means $\bar{\bm X}$ affect the performance of OBP, we introduced a bias term in the sample means. Specifically, for each area, the sample means were adjusted by setting: $\bar{\bm x} = \bar{\bm X} + \text{bias}$, where the bias term varied to take on negative, positive, and zero values.

\begin{figure}[h!]
    \centering
        \caption{Relationship between $|\bar X_i-\bar x_i|$ and ${\text{MSPE}_i}$ for area $i=1$}
    \includegraphics[scale=0.5]{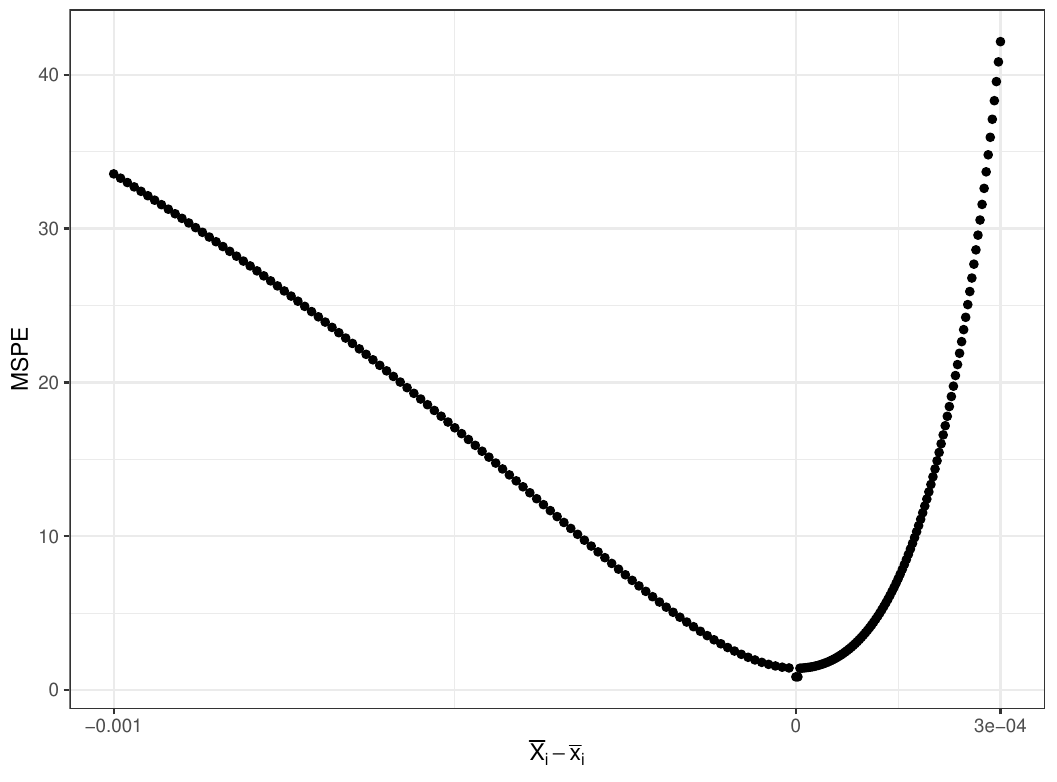}
    \label{fig:enter-label}
\end{figure}

The design-based MSPE for OBP using unit-level auxiliary variables ($\text{MSPE}_i$) was calculated for each area, and these values were compared with the magnitude of the absolute difference $|\bar{\bm X} - \bar{\bm x}|$. The results were visualized in Figure \ref{fig:enter-label}, which illustrates that the simulated ${\text{MSPE}_i}$ for OBP-UNIT increases with the difference between $\bar X_i$ and $\bar x_i$ for area $i=1$. Notably, when $\bar X_i - \bar x_i = 0$, corresponding to the unit-context model, simulated ${\text{MSPE}_i}$ is lower compared to other scenarios with non-zero differences. These results align with our theoretical design-based expectation of  $Q$ in section \ref{sec:2}.

\subsection*{Remark 2: Minimizing $Q(\bm{\psi})$ vs Minimizing $\rm{MSPE}(\tilde{\theta}(\bm{\psi}))$}
\label{remark2}
Another possible explanation for the underperformance of OBP-UNIT might be that the basic OBP equation, valid for fixed $\bm{\bm\psi}$, may not hold for random $\bm{\bm\psi}$, as in the case of BPE or MLE. To investigate this, we begin with the basic OBP equation: 
    \begin{equation}
        \rm{E}(|\tilde{\bm{\theta}}(\bm\psi) - \bm{\theta}|^2) = \rm{E}\{Q(\bm{\psi})\}.
        \label{MSPE=EQ2}
    \end{equation}
Let $\hat{\bm\psi}$ be the BPE of $\bm\psi$ obtained by minimizing the observed MSPE, and let $\tilde{\bm \psi}$ denote the MLE. The corresponding OBP-UNIT and EBLUP-UNIT are $\tilde{\bm\theta}(\hat{\bm\psi})$ and $\tilde{\bm\theta}(\tilde{\bm\psi})$, respectively. If the equation \eqref{MSPE=EQ2} holds for random $\bm\psi$, one would have 
\begin{equation}
    \rm{E}(|\tilde{\bm\theta}(\hat{\bm\psi}) - \bm\theta|^2) = \rm{E}\{Q(\hat{\bm\psi})\} \leq \rm{E}\{Q(\tilde{\bm\psi})\} = \rm{E}(|\tilde{\bm\theta}(\tilde{\bm\psi}) - \bm{\theta}|^2).
    \label{bpe_mle}
\end{equation}
The inequality follows from the definition of the BPE. To test this, we compare the four terms in the inequality using the same simulation settings as in previous analyses. The results are presented in Tables \ref{tab:4terms1} and \ref{tab:4terms2}.

When the assumed model is completely different from the true model, Table \ref{tab:4terms1} indicates that $\rm{E}(|\tilde{\bm\theta}(\hat{\bm\psi)} - \bm\theta|^2)$, the first  term of (\ref{bpe_mle}), could be much larger than $\rm{E}\{Q(\hat{\bm\psi})\}$, the second term of (\ref{bpe_mle}), unless $m$ is extremely large. But $\rm{E}\{Q(\tilde{\bm\psi})\}$, third term of (\ref{bpe_mle}) is approximately equal to $\rm{E}(|\tilde{\bm\theta}(\tilde{\bm\psi}) - \bm\theta|^2)$,  the third term of (\ref{bpe_mle}) even for smaller $m$. When the assumed model is partially correct, Table \ref{tab:4terms2} shows neither  $\rm{E}(|\tilde{\bm\theta}(\hat{\bm\psi}) - \bm\theta|^2) = \rm{E}\{Q(\hat{\bm\psi})\}$ nor $\rm{E}\{Q(\tilde{\bm\psi})\} = \rm{E}(|\tilde{\bm\theta}(\tilde{\bm\psi}) - \bm\theta|^2)$ of (\ref{bpe_mle})  holds, with $\rm{E}(|\tilde{\bm\theta}(\hat{\bm\psi}) - \bm\theta|^2)$ being much larger than $\rm{E}{Q(\hat{\bm\psi})}$, especially for smaller $m$ (e.g., 40).  Thus our simulation supports the argument that MSPE of OBP-UNIT could be larger than that of MSPE of EBLUP-UNIT unless $m$ is extremely and unreasonably large.

One potential factor contributing to these differences  between $\rm{E}(|\tilde{\bm\theta}(\hat{\bm\psi}) - \bm\theta|^2)$ and $\rm{E}\{Q(\hat{\bm\psi})\}$ is the numerical stability of $\hat{\bm \psi}$, BPE of $\bm\psi$. Particularly when $m = 40$, nearly 20\%, of BPE estimates yielding a zero value for the ratio $\sigma^2_u/\sigma^2_e$. The performance of OBP-UNIT appears to depend on value of $m$. The performance of OBP-UNIT improves for extremely large $m$ (e.g., 4000 or 40000), where both sides of the inequality of \eqref{bpe_mle},  i.e., $\rm{E}(|\tilde{\bm\theta}(\hat{\bm\psi}) - \bm\theta|^2) = \rm{E}\{Q(\hat{\bm\psi})\}$ and $\rm{E}\{Q(\tilde{\bm\psi})\} = \rm{E}(|\tilde{\bm\theta}(\tilde{\bm\psi}) - \bm\theta|^2)$, tend to be approximately hold, and thus OBP-UNIT eventually perform marginally better than EBLUP-UNIT under the model misspecifications. We hypothesize that the BPE of variance components stabilizes as $m$ increases, suggesting that future research should work on improving the BPE of variance components before proceeding the OBP method under unit-level models.
\begin{table}[h!]
    \centering
    \caption{Comparison among the four terms in \eqref{bpe_mle} under Table \ref{tab:table1} settings}
    \begin{tabular}{c|cccc}
    \toprule
    $(m, b)$ & $\rm{MSPE}(\tilde{\bm\theta}(\hat{\bm\psi}))$& $\rm{E}\{Q(\hat{\bm\psi})\}$ & $\rm{E}\{Q(\tilde{\bm\psi})\}$ & $\rm{MSPE}(\tilde{\bm\theta}(\tilde{\bm\psi}))$  \\
    \midrule
    (40, 5) & 1.775 & 0.448 & 1.479 & 1.474\\
    (100, 5)& 1.409 & 0.918 & 1.493 & 1.495 \\
    (400, 5)& 1.352 & 1.211 & 1.500 & 1.508 \\
    (1000, 5)& 1.384 & 1.323 & 1.504 & 1.513 \\
    (2000, 5)& 1.414 & 1.379 & 1.521 & 1.522 \\
    (4000, 5)& 1.414 & 1.398 & 1.512 & 1.515 \\
    \hline
    (40, 10) & 3.919 & -1.189 & 1.649 & 1.522\\
    (100, 10)& 2.364 & 0.493 & 1.551 & 1.513 \\
    (400, 10)& 1.711 & 1.231 & 1.508 & 1.510 \\
    (1000, 10)& 1.600 & 1.389 & 1.530 & 1.532 \\
    (2000, 10)& 1.509 & 1.412 & 1.500 & 1.496 \\
    (4000, 10)&1.485 &1.431 &1.493 &1.493\\
    \bottomrule
    \end{tabular}
    \label{tab:4terms1}
\end{table}
    \begin{table}[h!]
    \centering
    \caption{Comparison among the four terms in \eqref{bpe_mle} under Table \ref{tab:table2} settings}
    \begin{tabular}{c|cccc}
    \toprule
    $m$ & $\rm{MSPE}(\tilde{\bm\theta}(\hat{\bm\psi}))$& $\rm{E}\{Q(\hat{\bm\psi})\}$ & $\rm{E}\{Q(\tilde{\bm\psi})\}$ & $\rm{MSPE}(\tilde{\bm\theta}(\tilde{\bm\psi}))$  \\
    \midrule
    40 & 18.230 & -14.036 & 3.247 & 1.532\\
    100 & 7.958 & -4.809 & 1.586 & 1.514 \\
    400& 3.122 & -0.584 & 1.583 & 1.508 \\
    1000& 2.173 & 0.828 & 1.580 & 1.569\\
    2000&1.815 &1.131 & 1.546 &1.501\\
    4000&1.650 &1.267 &1.496 &1.505\\
    40000 &1.488 & 1.408 &1.462 &1.503\\
    \bottomrule
    \end{tabular}
    \label{tab:4terms2}
\end{table}

\section{Conclusion}\label{sec:4}
In this paper, we investigate the effects of misspecified mean structure and sampling variance in the well-known nested error regression  model on EBLUP and OBP that uses (i) unit-level auxiliary variables only and (ii)  area-level auxiliary variables only (i.e., unit context model). Through a series of simulations, we demonstrate that under significant model misspecifications, the OBP procedure for the area-level model and the unit context model outperforms the corresponding unit-level model that relies solely on unit-specific auxiliary variables.

We propose two potential reasons for the underperformance of OBP-UNIT: first, the difference between the sample and population means of the auxiliary variables may negatively affect the performance of OBP-UNIT; second, the numerical stability of the BPE for variance components using unit-level auxiliary variables could be a contributing factor. Our results suggest that utilizing the OBP procedure with area-level auxiliary variables is a promising alternative, particularly when challenges such as the lack of census information and model misspecification is a concern. 

Furthermore, when choosing between the unit-context model and the area-level model, we recommend the area-level model as it may offer greater stability. In the area-level model, the variance of sampling errors can be estimated using smoothing methods and is less dependent on the assumed model which may be misspecified. Finally, we recommend further theoretical and empirical research to explore and better understand these unexpected outcomes in the OBP for nested error models that utilize only unit-level auxiliary variables.
\bibliographystyle{chicago}
\bibliography{ref.bib}

\end{document}